\documentclass[10pt,aps,twocolumn,prb,superscriptaddress]{revtex4-2}

\usepackage{amsmath,amssymb,amsthm,mathrsfs,amsfonts,dsfont,amstext} 
\usepackage{textcomp,pbox}
\usepackage[export]{adjustbox}
\usepackage{soul} 
\usepackage{bm,xspace}
\usepackage{dcolumn,booktabs,url}
\usepackage[scaled]{helvet}
\usepackage{sansmath,gensymb}
\usepackage{tikz,graphicx,transparent,color}
\usepackage{multirow}
\usepackage[separate-uncertainty = true]{siunitx}
\usepackage{comment}
\usepackage{physics}
\usepackage{pdfpages}
\usepackage{array}
\usepackage{makecell}
\usepackage{tabularx}
\usepackage{float}

\newcommand{\extfig}[1]{Extended Data Fig. \ref{#1}}
\newcommand{\exttab}[1]{Extended Data Table. \ref{#1}}
\newcommand{\Rmnum}[1]{\uppercase\expandafter{\romannumeral #1}}

\newcolumntype{L}{>{\raggedright\arraybackslash}X}

\makeatletter
\AtBeginDocument{\let\LS@rot\@undefined}
\makeatother

\newcolumntype{C}[1]{>{\middleing\let\newline\\\arraybackslash\hspace{0pt}}m{#1}}

\usepackage[colorlinks=true]{hyperref}
\usepackage{graphicx}

\hypersetup{
     colorlinks   = true,
     citecolor    = blue,
     linkcolor    = blue,
     urlcolor     = blue     
}

\newcommand{\ee}{\mathrm{e}}
\newcommand{\ii}{\mathrm{i}}

\newcommand{\ketup}{{\ket{\uparrow}}}
\newcommand{\ketdown}{{\ket{\downarrow}}}

\makeatletter
\def\maketitle{
\@author@finish
\title@column\titleblock@produce
\suppressfloats[t]}
\makeatother

\newcommand{\lqcc}{Laboratory of Quantum Information, University of Science and Technology of China, Hefei 230026, China}
\newcommand{\anhuikey}{Anhui Province Key Laboratory of Quantum Network, University of Science and Technology of China, Hefei 230026, China}
\newcommand{\cascenter}{CAS Center For Excellence in Quantum Information and Quantum Physics, University of Science and Technology of China, Hefei 230026, China}
\newcommand{\hfnl}{Hefei National Laboratory, University of Science and Technology of China, Hefei 230088, China}

\begin{document}

\title{Bell nonlocality with directly generated telecom-band spin--photon entanglement}

\author{Dong-Yu Huang}
\thanks{D.-Y. H. and J. W. contributed equally to this work.}
\affiliation{\lqcc}
\affiliation{\anhuikey}
\affiliation{\cascenter}
\affiliation{\hfnl}

\author{Jian Wang}
\email{jwang28@ustc.edu.cn}
\affiliation{\lqcc}
\affiliation{\anhuikey}
\affiliation{\cascenter}

\author{Xiao-Long Zhou} 
\affiliation{\lqcc}
\affiliation{\anhuikey}
\affiliation{\cascenter}

\author{Ze-Min Shen}
\affiliation{\lqcc}
\affiliation{\anhuikey}
\affiliation{\cascenter}

\author{Si-Jian He}
\affiliation{\lqcc}
\affiliation{\anhuikey}
\affiliation{\cascenter}
\affiliation{\hfnl}

\author{Qi-Yang Huang}
\affiliation{\lqcc}
\affiliation{\anhuikey}
\affiliation{\cascenter}

\author{Yi-Jia Liu}
\affiliation{\lqcc}
\affiliation{\anhuikey}
\affiliation{\cascenter}
\affiliation{\hfnl}

\author{Yu-Shu Chen}
\affiliation{\lqcc}
\affiliation{\anhuikey}
\affiliation{\cascenter}

\author{Quan Jiang}
\affiliation{\lqcc}
\affiliation{\anhuikey}
\affiliation{\cascenter}

\author{Chuan-Feng Li}
\email{cfli@ustc.edu.cn}
\affiliation{\lqcc}
\affiliation{\anhuikey}
\affiliation{\cascenter}
\affiliation{\hfnl}

\author{Guang-Can Guo}
\affiliation{\lqcc}
\affiliation{\anhuikey}
\affiliation{\cascenter}
\affiliation{\hfnl}

\begin{abstract}
Quantum nonlocality, typically revealed through entanglement distribution across quantum networks, is a cornerstone of quantum information science. Long-distance distribution of entanglement requires the information carrier, i.e. flying photons, to operate in the minimum-loss telecom band of optical fiber. While extensive efforts have been devoted to the direct generation of entanglement between C-band telecom photons and various stationary spins, the verification of quantum nonlocality remains an outstanding challenge. Here, utilizing a dipole transition in rubidium atoms with a wavelength of 1530 nm and a cavity-assisted protocol, we achieve resonant excitation and direct emission of C-band telecom photons from a single atom, generating spin--photon entanglement with a measured Bell state fidelity exceeding 91.4\%. We then verify Bell nonlocality by observing a Bell inequality violation of $2.455(77) > 2$ using this high-quality entangled pair. These results extend the wavelength of a single-atom quantum emitter to the telecom C-band, achieving sufficiently high-fidelity spin--photon entanglement to finally verify Bell nonlocality. This work thereby provides a promising building block for a large-scale atom-based quantum network capable of distributed quantum metrology and long-distance quantum communication.
\end{abstract}

\maketitle


As a cornerstone of quantum information science \cite{2014RMPBellNonlo}, quantum nonlocality plays a key role in various applications of quantum networks, including distributed quantum computing and quantum communication whose security is guaranteed by physical laws \cite{2022rmpsecurity}. These technologies require the distribution of entanglement between different quantum nodes, which typically relies on establishing entanglement between a flying photon and a local stationary spin \cite{kimble2008}. Limited by the no-cloning theorem \cite{noclone}, photonic qubits are extremely vulnerable to the exponential loss during transmission \cite{cz1998repeater}. Therefore, the operation of a practical quantum network is more sensitive to wavelength-dependent loss than its classical counterparts which allow amplification of optical signals among the general telecom bands. As a result, the practical deployment of quantum networks demands minimal photon transmission loss—a requirement met by the telecom C-band \cite{2023nrpphoton, bands_knowledge}, which offers extremely low attenuation (below 0.16 dB/km) to support thousand-kilometer-scale repeaterless quantum communication \cite{wangshuangqkd, panqkd} and excellent compatibility with existing fiber infrastructure.

To date, extensive efforts have been devoted to the utilization of quantum systems possessing optical transitions in the telecom C-band \cite{riedmatten_er_purcell, zhouerstorage, er_reiserer_stable, er_indis_thomp, er_hxtang,  qdtelecomreview, qd_telecom_htoon,er_groeblacher, qd_indis_syperek, qd_indis_hoefling, qd_teleemission_shields}. Despite the achievements in spin--photon entanglement generated by these emitters \cite{2024qdentangle, 2025erentangle}, the high-fidelity entanglement sufficient for verifying quantum nonlocality remains a significant challenge due to the limited coherence. Benefiting from the excellent coherence and well-established state manipulation \cite{rempe2015rmp, bernien2023npj, saffmanqc, lukin2024qec}, neutral atoms offer promising prospects for quantum networking. Single ytterbium atoms have been utilized for direct generation of telecom-band atom--photon entanglement with a raw fidelity of 90\% \cite{2025ybentangle}, yet the transition lies in E-band where fiber attenuation is significantly higher \cite{fund_photonics}. Other neutral atoms, e.g., rubidium, have also been explored for long-distance entanglement distribution via quantum frequency conversion (QFC) to the telecom S-band \cite{weinfurter_qfc_2020,33km,weinfurter_2026_metropolitan} and O-band \cite{pandiqkd}. However, the associated noise poses obstacles to interfacing with the lowest-loss telecom C-band for quantum communication, while finite conversion efficiency limits the deployment of a large-scale quantum network.

In this work, we present the first direct interface between a neutral atom and the telecom C-band using a dipole transition in rubidium \cite{kuzmich_telecom, duan1530}. Following the fast two-photon resonant excitation of the single atom, we observe correlated two-photon emission from the cascaded transition. A cavity-assisted transfer scheme, combined with photon heralding \cite{rempe_xcav_776}, then generates entanglement between the C-band photon polarization and the atomic ground-state spin. A Bell state fidelity with a lower bound of 91.4\% is characterized by correlation measurement of the entangled pair, which relies on coherent operation and cavity-assisted state readout of the atom \cite{2025readout}. Finally, the violation of Bell inequality $2.455(77) > 2$ is observed, manifesting the Bell nonlocality within the entangled pair.

\vspace{-0.1cm}
\section*{Experimental protocol and telecom photon characterization}
\vspace{-0.3cm}

\begin{figure}[h]
    \centering
    \includegraphics[width=0.45\textwidth]{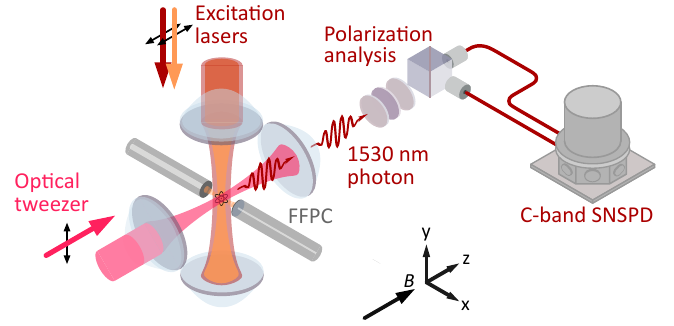}
    \caption{
    Schematic diagram of the experimental setup. An optical tweezer traps the atom at the center of the FFPC. A magnetic field of 170 mG is applied along the $z$ direction as a quantization axis. The emitted telecom photons are collected by one of the aspherical lenses placed along the quantization axis. For entanglement characterization, they are directed through a subsequent polarization analysis module and are detected by a two-channel superconducting nanowire single photon detector (SNSPD). The spectral filtering setup of the photons is omitted for clarity.} 
    \label{Fig1}
\end{figure}

Our experimental system consists of a single rubidium-87 atom trapped at the focus of four aspherical lenses, one of which serves as the collection channel of the emitted 1530 nm photons as shown in Fig. \ref{Fig1}. The atom is also placed at the center of a fiber-based Fabry-Pérot microcavity (FFPC) \cite{reichel_cavity_bec, rempexcav, lukinfibercav, 2025readout} for entanglement generation and atomic state readout. Different from our previous work \cite{xiaolong2photon}, the collection of the telecom photons requires precise and stable positioning of the atom at the focal plane of the lens. Therefore, instead of an optical lattice, we apply an optical tweezer with an 850 nm laser beam co-propagating with the collection beam.

\begin{figure*}[]
    \centering
    \includegraphics[width=0.9\textwidth]{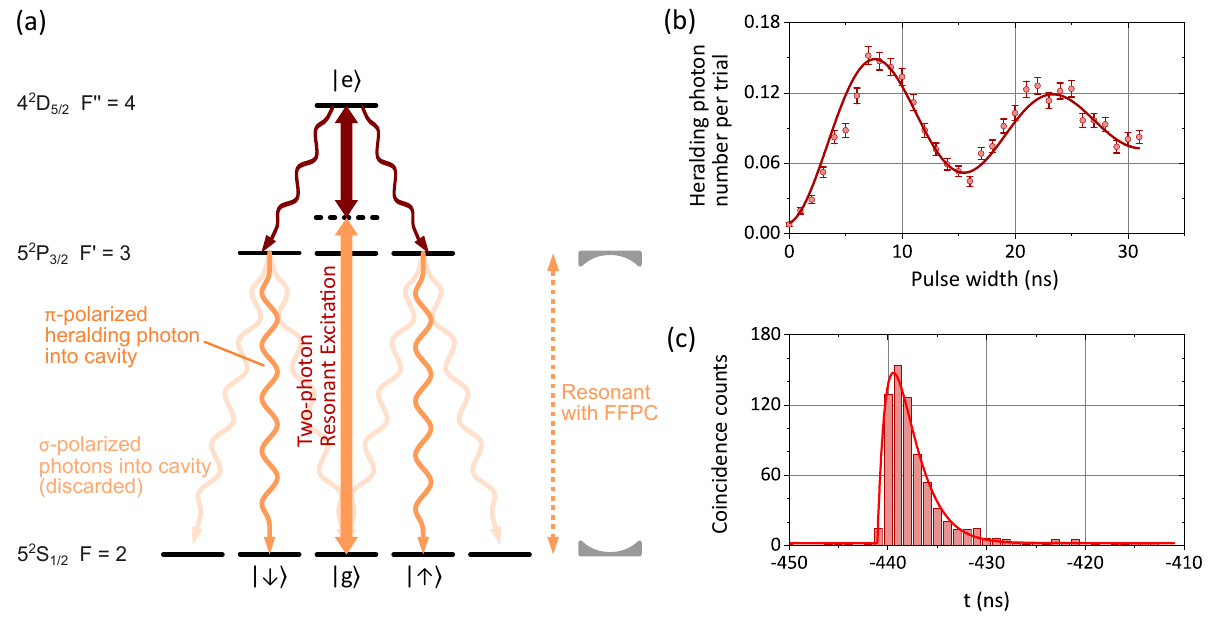}
    \caption{(a) Protocol for entanglement generation between a single $^{87}\text{Rb}$ atom and a C-band photon. Not all the Zeeman states are shown. (b) Rabi oscillation from $\ket{g}$ to $\ket{e}$. The relative population of $\ket{e}$ is indicated by the detection of 780 nm heralding photons. A Rabi frequency of 61 MHz is extracted. (c) Two-photon coincidences between the 1530 nm telecom photons and the 780 nm heralding photons over a 10-hour experimental run. The horizontal axis shows the arrival time difference between the 780 nm photon and the 1530 nm detection event, with the large offset arising from the propagation delay of the 1530 nm signal through the long fiber and coaxial cable to the SNSPD located in another room. The width of the coincidence represents the Purcell-enhanced lifetime of the 780 nm photon which is 1.4 ns here, indicating an atom-cavity cooperativity of 9.}
    \label{Fig2}
\end{figure*}

The protocol for the direct generation of spin--photon entanglement is illustrated in Fig. \ref{Fig2} (a), which utilizes the cascaded transition $4D$--$5P$--$5S$ in rubidium with the upper transition lying in the telecom C-band. The atom is first initialized in a ground state $\ket{g} := \ket{5^2S_{1/2},F=2,m=0}$ by optical pumping. A two-photon excitation pulse drives the atom to the excited state $\ket{e} := \ket{4^2D_{5/2},F''=4,m=0}$ followed by a spontaneous decay to the intermediate level $\ket{5^2P_{3/2},F'=3,m=0,\pm1}$. As the axis of the collection lens is placed along the quantization axis defined by a static magnetic field, telecom photons from the $\pi$ decay channel cannot be collected \cite{wein2012entangle, 33km, saffman_parabolic_prxq2026, sup}, resulting in the following entangled state of the collected photonic polarization qubit and the atomic spin:
\begin{align}
    \ket{\Psi'} = \frac{\ket{\uparrow',\sigma_-}+ \ket{\downarrow',\sigma_+}}{\sqrt{2}},
\end{align}
where
\begin{align}
\ket{\uparrow'}&:=\ket{5^2P_{3/2},F=3,m=+1}\\
\ket{\downarrow'}&:=\ket{5^2P_{3/2},F=3,m=-1}. 
\end{align}
Then shortly a Purcell-enhanced decay brings the atom from the intermediate level back to the long-lived ground level, with a 780 nm $\pi$-polarized photon heralding the successful creation of a final entangled state
\begin{equation}
   \ket{\Psi} = \frac{\ket{\uparrow,\sigma_-}+ \ket{\downarrow,\sigma_+}}{\sqrt{2}}. 
\end{equation}
where
\begin{align}
\ket{\uparrow}&:=\ket{5^2S_{1/2},F=2,m=+1}\\
\ket{\downarrow}&:=\ket{5^2S_{1/2},F=2,m=-1}. 
\end{align}

Here we apply a polarization prism at the out-coupling port of the FFPC to selectively collect the 780 nm $\pi$-photons with the $\sigma$-photons being discarded. This is not physically essential, since a polarization-nondegenerate FFPC can enhance the $\pi$ transition only \cite{reichel_nondegenerate, zhoukun_nondegenerate, rempexcav}, which has been utilized for heralded generation of atom--photon entanglement in another cascaded transition \cite{rempe_xcav_776}.

The Rabi oscillation between $\ket{g}$ and $\ket{e}$ is shown in Fig. \ref{Fig2} (b). The two-photon excitation pulse consists of a 780 nm laser beam chopped by an acousto-optical modulator (AOM) and a 1530 nm laser chopped by an AOM and an electro-optic intensity modulator. A single-photon detuning is chosen to be 1 GHz to avoid scattering through the intermediate states. Note that a fast excitation can also populate a nearby excited state $\ket{e'} := \ket{4^2D_{5/2},F''=2,m=0}$ which is undesired in the protocol above. To avoid this, a two-photon Rabi frequency $\Omega = 60$ MHz is determined such that the effective Rabi frequency $\Omega'$ between $\ket{g}$ and $\ket{e'}$ is close to $2\Omega$ to minimize the undesired population in $\ket{e'}$ (Methods). Once the atom is in the excited state $\ket{e}$, it decays to the ground state through the cascaded channel with two photons, within which the 780 nm photon is collected and out-coupled by the FFPC with a high probability, indicating a successful excitation.

The collection of the heralding 780 nm photon offers a chance of detecting the telecom photon with a high signal-to-noise ratio through two-photon coincidence. As shown in Fig. \ref{Fig2} (c), after a $\pi$-pulse excitation, we record the temporal correlation of the photon counts from photon detectors for the telecom and 780 nm photon respectively. The Purcell-enhanced decay of the lower transition gives a sharp correlation spike, which serves as the detection signal of the telecom photon and completes the entanglement generation protocol. The overall heralded detection efficiency of a 1530 nm photon is $5 \times 10^{-5}$, which is the product of the 780 nm heralding efficiency $0.1$ and the detection efficiency of the 1530 nm photon $5\times 10^{-4}$ \cite{sup}. This relatively low efficiency mainly arises from the aberration of the aspherical lens ($10^{-3}$) for free-space collection, with the remainder (0.5) coming from the transmission losses of the optical components, diffraction by a grating filter, fiber coupling and the detection efficiency of the SNSPD.

\begin{figure*}[t]
    \centering
    \includegraphics[width=\textwidth]{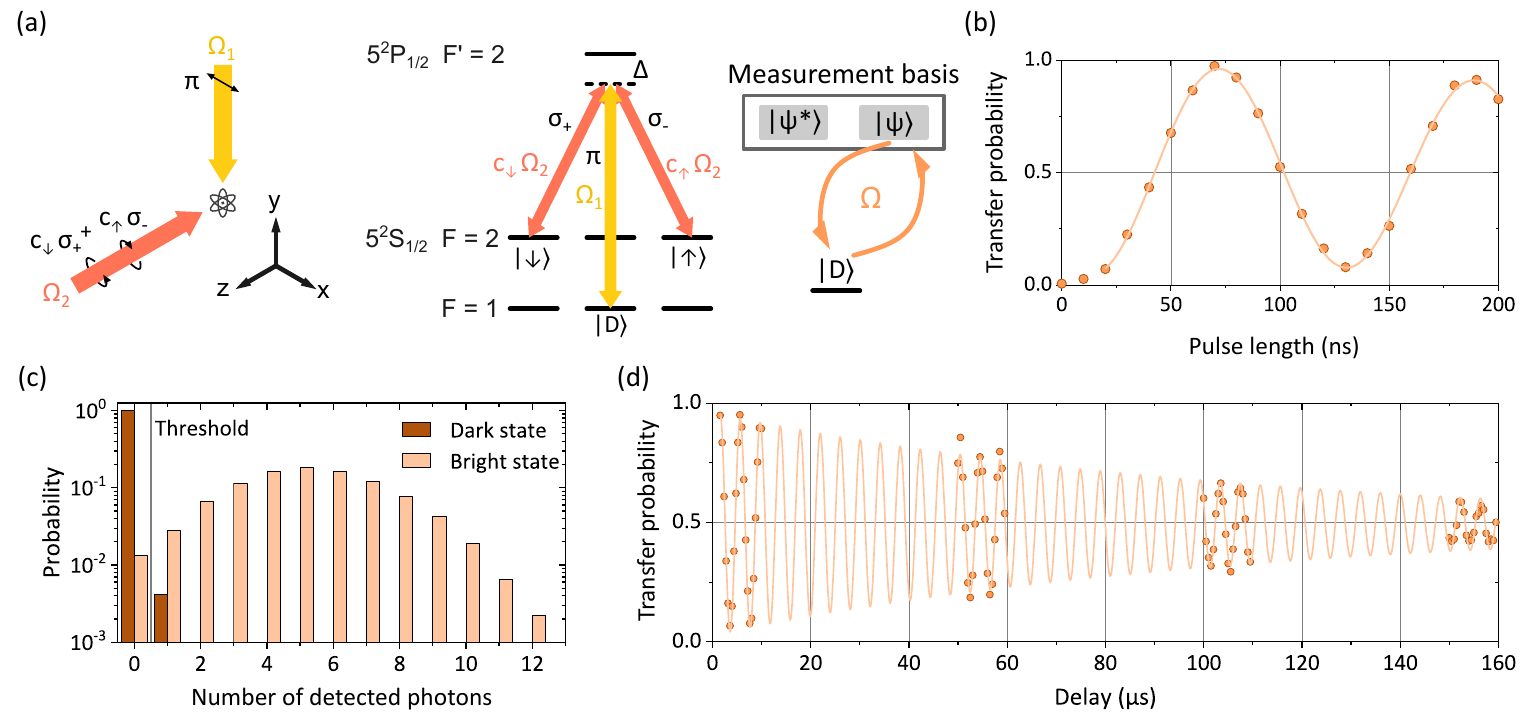}
    \caption{(a) State-selective transfer scheme of atomic qubit measurement. For a laser configuration with the polarization of $\Omega_2$ expressed as $c_\downarrow \ket{\sigma_+} + c_\uparrow\ket{\sigma_-}$, only the superposition state $\ket{\psi} := c_\downarrow\ketdown + c_\uparrow\ketup$ is coupled to $\ket{D}$, while its orthogonal opponent $\ket{\psi^*}$ remains unaffected. Particularly, when $c_\downarrow = 0$, i.e. the polarization of the laser field $\Omega_2$ is $\sigma_-$, the configuration leads to the measurement of eigenstates $\{ \ket{\uparrow}, \ket{\downarrow}\}$. Not all the Zeeman states are shown in the diagram. (b) Rabi oscillation between $\ket{\leftarrow} := (\ket{\downarrow}+ \ket{\uparrow})/\sqrt{2}$ and $\ket{D}$. (c) Probability distribution of the detected photon number during the readout. An 800-ns readout pulse is applied to scatter enough photons into the FFPC. The dark state and bright state refer to the hyperfine levels $\ket{5^2S_{1/2},F=1}$ and $\ket{5^2S_{1/2},F=2}$, respectively. (d) Coherence characterization of the atomic qubit. As the relative phase accumulates between the two components of $\ket{\leftarrow}$, the population in $\ket{\leftarrow}$ varies over time, leading to an oscillating transfer probability to $\ket{D}$. An exponential fit indicates an $1/\ee$ contrast decay time of 107 $\mu$s.}
    \label{Fig3}
\end{figure*}

\section*{Measurement of the atomic qubit}

The full characterization of spin--photon entanglement requires the measurement of both qubits in various bases. As the measurement of photonic polarization by waveplates and polarization prisms is relatively straightforward, the measurement of the atomic qubit demands projection of the atom onto the eigenstates $\{\ketdown,\ketup\}$ or their superpositions. Since both states lie in the $F=2$ manifold, the general idea here is to transform the projective measurement of the qubit to the discrimination of atomic hyperfine levels $F=1$ and $F=2$, hence a cavity-assisted fluorescence-based state readout can be applied \cite{2025readout}. Therefore, suppose that a measurement is carried out in an arbitrary basis $\{ \ket{\psi},\ket{\psi^*} \}$, where
\begin{align}
    \ket{\psi} &= c_\downarrow\ketdown + c_\uparrow\ketup \notag \\
    \ket{\psi^*} &= c_\uparrow^*\ketdown - c_\downarrow^*\ketup \label{definePsi}
\end{align}
so that $\braket{\psi}{\psi^*}=0$. Then we need to transfer the population on $\ket{\psi}$ to the $F=1$ level which is the auxiliary state $\ket{D}:=\ket{F=1,m_F=0}$ specifically, while the population on $\ket{\psi^*}$ remains unaffected in the $F=2$ level. Since $\ket{D}$ is a dark state in the subsequent fluorescence readout \cite{2025readout}, a bright signal indicates that the atom was previously in $\ket{\psi^*}$, thereby completing the measurement in this basis.

To achieve this transfer, inspired by previous work in state-selective ionization \cite{weinionization}, we perform state-selective transfer using stimulated Raman transition to address an arbitrary measurement basis. The stimulated Raman transition is illustrated in Fig. \ref{Fig3} (a) where the polarization of Raman laser $\Omega_2$ determines the state to be transferred.

Two lasers with a frequency separation of 6.8 GHz are phase-locked to a radio-frequency reference (Methods), and are focused on the atom along ($\sigma$) and perpendicular to ($\pi$) the quantization axis respectively. The single-photon detuning from the $5^2P_{1/2}$ level is 40 GHz. In order to calibrate the state selective transfer, we reverse the process by optically pumping the atom to the state $\ket{D}$ which is the dark state in the atomic state readout process, and transfer the atom to an equal superposition of $\ket{\uparrow}$ and $\ket{\downarrow}$ by setting the laser $\Omega_2$ as linearly polarized. The transfer probability reaches 97\% and a separate characterization of the cavity-assisted fluorescence readout shows a fidelity of 99.1\% as shown in Fig. \ref{Fig3} (c). The transfer pulse lasts for 70 ns, which is sufficiently short compared to the Larmor precession period between state $\ket{\downarrow}$ and $\ket{\uparrow}$, such that the population of the target superposition is not leaked to the orthogonal state during the transfer.

One major cause of errors in the distribution of spin--photon entanglement is the decoherence of atomic qubits during the protocol \cite{33km, pandiqkd}. With the full toolkit for atomic state analysis, we can characterize the coherence property of the atomic qubit by observing the evolution of the superposition state $\ket{\leftarrow}:=(\ket{\uparrow}+\ket{\downarrow})/2$. Similar to a Ramsey sequence in principle, the process is carried out as follows. The atom is first prepared at ground state $\ket{D}$ and then transferred to $\ket{\leftarrow}$ with a Raman pulse. The two components of $\ket{\leftarrow}$ gradually acquire a relative phase by Larmor precession in the magnetic field along the quantization axis. After a certain period of time, another transfer is performed with the laser configuration identical to the prior one. As shown in Fig. \ref{Fig3} (d), the transfer probability is determined by the atomic state which shows an oscillation with a frequency twice as the Larmor frequency $\omega_\mathrm{L} = 120$ kHz. An $1/e$ coherence time of the qubit is 107 $\mu$s as shown by the oscillation curve. This corresponds to a magnetic field fluctuation of approximately $1$ mG, which may result from the insufficient compensation of the magnetic
noise at the position of the atom.

\begin{figure*}[t]
    \centering
    \includegraphics[width=\textwidth]{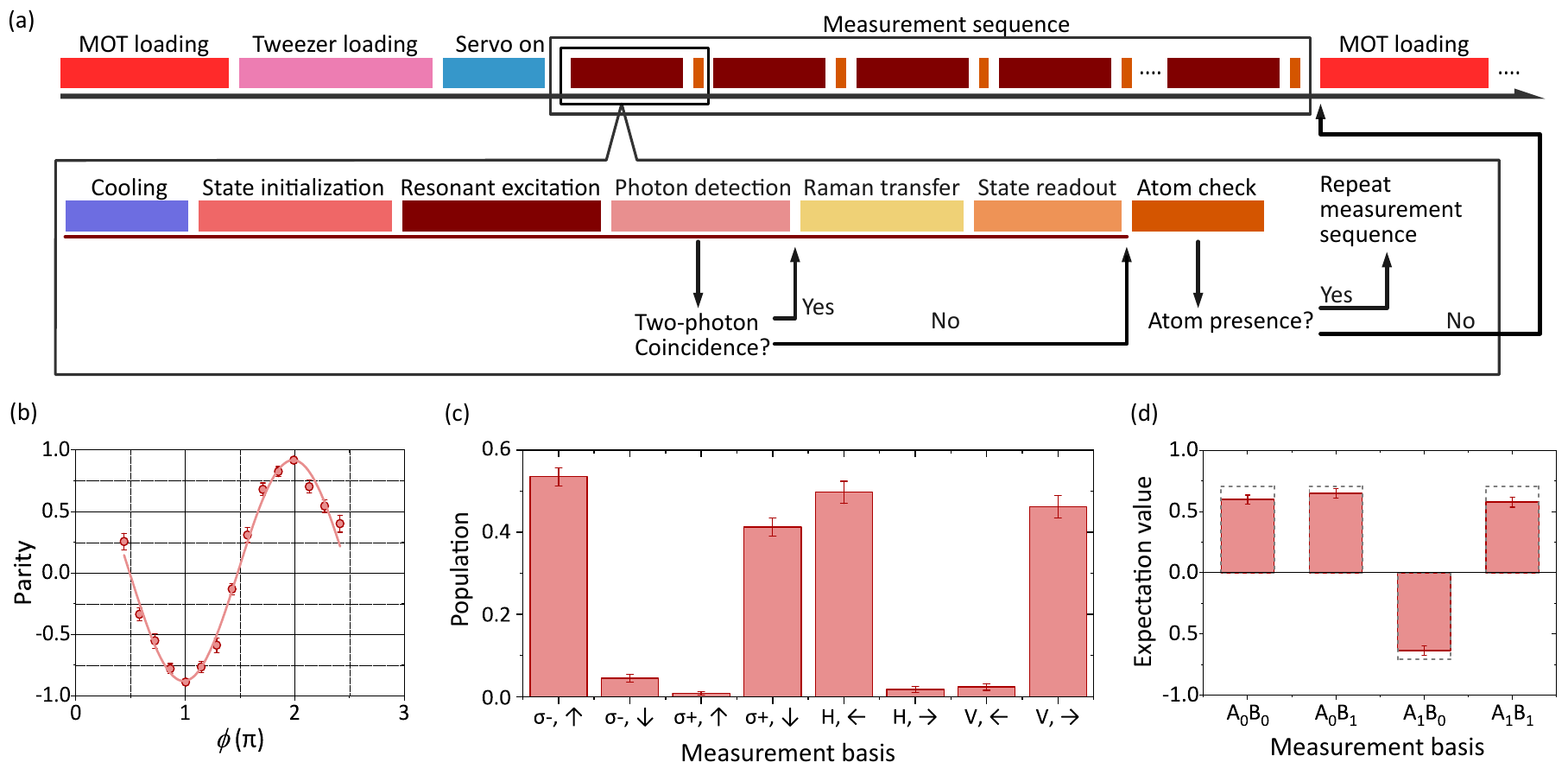}
    \caption{(a) Experimental time sequence for characterizing the generated spin--photon entanglement. The main sequence begins with a 500-$\mu$s laser cooling and 40-$\mu$s state initialization. The atom is then excited to $\ket{e}$ and an FPGA controller starts to record two-photon coincidence. Once a coincidence event occurs, the atom is measured by Raman transfer and cavity-assisted fluorescence readout. Each run is ended with a fluorescence check of the presence of the atom, whose loss ends the main sequence and starts MOT loading again. (b) Nonclassical parity oscillation of the entangled state. The measurement basis of the photon is set to be the linear polarization, while the atomic basis is varied by changing the time of Larmor precession followed by a projective measurement onto $\ket{\leftarrow}$. (c) Measured population on various bases for characterization of Bell state fidelity. (d) Measured correlations under different bases for testing Bell inequality. A CHSH parameter of 2.455(77) is measured which verifies Bell nonlocality. The same expectations of an ideal Bell state are plotted with dashed columns.}
    \label{Fig4}
\end{figure*}

\section*{Characterization of the spin--photon entanglement}

The characterization of spin--photon entanglement is carried out with a time sequence shown in Fig. \ref{Fig4} (a). After a single atom is loaded into the optical tweezer, a digital servo is triggered on to stabilize the magnetic field and powers of the Raman lasers (Methods). After laser cooling and state initialization, the atom is excited and an FPGA-based controller records the measurement outcome of the 1530 nm photon by detecting two photon coincidence. Then the atomic state is measured by the transfer-readout process as described above.

The nonclassical correlation of the entangled state is characterized by observing the parity oscillation as shown in Fig. \ref{Fig4} (b). Here, the photonic polarization measurement basis is $\{ \ket{H},\ket{V} \}$, and the Raman lasers are in the same configuration as the one for the transfer of $\ket{\leftarrow}$. The phase of the atomic measurement basis is varied by adding a delay $\tau$ prior to the Raman transfer pulse, such that the projection measurement onto any superposition $(\ket{\uparrow}+\ee^{\ii\phi}\ket{\downarrow})/\sqrt{2}$ can be realized, where $\phi = 2\omega_\mathrm{L} \tau$. An oscillation contrast of 0.90 is observed, revealing the nonclassical correlation of the entangled state.

The fidelity of the entangled state is estimated with the same approach applied in \cite{monroe2004} and detailed in \cite{sup}, which is performed by measuring the population of the state in both classical and nonclassical bases. In our case, the lower bound of the Bell state fidelity  is given by 
\begin{align}
    F > & \frac{1}{2} \Big(
P_{\sigma_-\uparrow} + P_{\sigma_+\downarrow} - 2\sqrt{P_{\sigma_-\downarrow} \;P_{\sigma_+\uparrow}} \notag \\
\;&  +P_{H\leftarrow} + P_{V\rightarrow} - P_{H\rightarrow} - P_{V\leftarrow}
\Big),
\end{align} 
where $P_{ab} := \rho_{ab,ab}$ denotes the population of the state with photonic polarization in $\ket{a}$ and atomic qubit in $\ket{b}$. The measurement on the classical basis is performed by setting both the 1530 nm polarization analysis system and the Raman laser $\Omega_2$ to circularly polarized configuration. Based on the result shown in Fig. \ref{Fig4} (c), a lower bound of the fidelity is evaluated as 91.4\%, which significantly surpasses the requirement for a nonlocality Bell test \cite{hansonloophole}. The infidelities mainly originate from state initialization ($\sim2\%$), polarization error in heralding ($\sim 3\%$) and insufficient Raman transfer in atomic state measurement ($3\%$).

The verification of Bell nonlocality can be performed by observing the violation of a Clauser-Horne-Shimony-Holt (CHSH) type Bell inequality \cite{xiaolong2photon} to rule out any local hidden variable models. In our experiment, the test requires the measurement bases of the 1530 nm photon and the atom to be set at a certain configuration to observe the maximum violation. Here, the two photonic bases are both set to be linear polarizations as $A_0 = \{ \ket{H},\ket{V} \}$ and $A_1 = \{ \ket{A},\ket{D} \}$ where the polarization of $A_1$ is $45\degree$ away from $A_0$. For the atomic qubit, we choose the basis $B_n$ as the projection onto state $\ket{\phi_n} = \ket{\uparrow} + \ee^{\ii\phi_n}\ket{\downarrow}$ where $\phi_0 = 3\pi/4$ and $\phi_1 = 5\pi/4$. By assigning two different measurement outcomes of a basis with values $\pm 1$, we obtain the expectations in four measurement configurations as shown in Fig. \ref{Fig4} (d). The CHSH parameter is calculated as
\begin{align}
    S &= |E(A_0B_0) + E(A_0B_1) - E(A_1B_0) + E(A_1B_1)| \\
    & = 2.455(77)>2,\notag
\end{align} 
which shows a definitive violation of the inequality with 5.9 standard deviations, manifesting Bell nonlocality with the generated spin--photon entangled pair.

\section*{Conclusion and Discussion}

In this work, we generate spin--photon entanglement in the telecom C-band with a direct emission protocol, extending the operational wavelength of single-atom emitters to the spectral range of minimal loss in optical fiber. Nonclassical correlation of the entangled state is observed and a Bell state fidelity over 91.4\% is measured. Furthermore, this entangled pair shows a violation of Bell inequality $2.455(77) > 2$ for the first time, demonstrating Bell nonlocality which is an essential resource for various quantum communication protocols, providing a promising building block toward practical application of telecom-band quantum emitters. 

We note that the generation efficiency in this work cannot match previous demonstrations with quantum frequency conversion \cite{33km} and direct E-band emission \cite{2025ybentangle}, which mainly results from the inefficient free-space collection in our setup \cite{sup}. This can be significantly improved by more than an order of magnitude using microscope objectives specially designed for C-band, matching state-of-the-art performance in free-space atom--photon entanglement \cite{33km, 2025ybentangle, pandiqkd}.

Further improvements can be made by employing a polarization-nondegenerate optical cavity \cite{reichel_nondegenerate, zhoukun_nondegenerate, rempexcav, rempe_xcav_776} to selectively enhance the $\pi$-decay channel only, which dispenses with heralding, resulting in higher brightness and overall entanglement generation efficiency \cite{sup}. The coherence time of atomic qubits can be extended to the scale of hundreds of milliseconds by encoding in magnetic-insensitive energy levels and dynamic decoupling \cite{rempe100ms, pandiqkd}, which makes it feasible for large-scale quantum communication. As proposed in \cite{rempetelepropos, bernientelepropos, coveytelepropos, covey_multiplexing_cavity}, an optical cavity resonant with the telecom transition can largely boost the collection efficiency of the telecom photons. The bandwidth of a single emitter can also be enhanced to more than tens of MHz by utilizing the Purcell regime in a microcavity \cite{2025readout}. 

The high optical access of fiber microcavities offers an opportunity for operating an intra-cavity atom array of tens of atoms using high-NA optical tweezers \cite{lukinfibercav}. For integration with larger-scale atom arrays, the zone-based architecture for neutral atom quantum computing \cite{endres6100, lukin_2026_faulttolerant} can be adopted. In this scheme, the cavity serves as an entanglement distribution region, with part of the atoms dynamically transported in and out for coupling. Furthermore, Rydberg-mediated gates can be made compatible with microcavity systems through fiber metallization \cite{northup_ion_ffpc_2023, cui_fiber_metalize_2026}, conductive coatings \cite{britton_zno_cavity}, and shielding of stray fields from piezoelectric components \cite{kong2018thesis_waterloo, chen2022thesis_rydber_ring}. This would further enable the implementation of protocols such as quantum repeaters and distributed quantum computing.

Based on the demonstration and feasible improvements of this work, one can expect a quantum network with functional quantum information processors directly interfacing the telecom C-band and fiber communication infrastructure, harnessing the full capability of a quantum internet \cite{2014clocknetwork, covey_clock_network, bernien2023npj, hanson2018review}.
\newline

\noindent
\textbf{Data availability}

\noindent
Data that support the findings of this study are available, upon reasonable request, from the corresponding authors.
\newline

\noindent
\textbf{Code availability} 

\noindent
The code used for data analysis during this study is available, upon reasonable request, from the corresponding authors.

\bibliography{ref}

@article{2014RMPBellNonlo,
  title = {Bell nonlocality},
  author = {Brunner, Nicolas and Cavalcanti, Daniel and Pironio, Stefano and Scarani, Valerio and Wehner, Stephanie},
  journal = {Rev. Mod. Phys.},
  volume = {86},
  issue = {2},
  pages = {419--478},
  numpages = {60},
  year = {2014},
  month = {Apr},
  publisher = {American Physical Society},
  doi = {10.1103/RevModPhys.86.419},
  url = {https://link.aps.org/doi/10.1103/RevModPhys.86.419}
}

@article{2022RMPSecurity,
  title = {Security in quantum cryptography},
  author = {Portmann, Christopher and Renner, Renato},
  journal = {Rev. Mod. Phys.},
  volume = {94},
  issue = {2},
  pages = {025008},
  numpages = {56},
  year = {2022},
  month = {Jun},
  publisher = {American Physical Society},
  doi = {10.1103/RevModPhys.94.025008},
  url = {https://link.aps.org/doi/10.1103/RevModPhys.94.025008}
}

@Article{kimble2008,
author={Kimble, H. J.},
title={The quantum internet},
journal={Nature},
year={2008},
month={Jun},
day={01},
volume={453},
number={7198},
pages={1023-1030},
issn={1476-4687},
doi={10.1038/nature07127},
url={https://doi.org/10.1038/nature07127},
}

@article{noclone,
  author = {Wootters, W. K. and Zurek, W. H.},
  title = {A Single Quantum Cannot Be Cloned},
  journal = {Nature},
  volume = {299},
  number = {5886},
  pages = {802-803},
  year = {1982},
  url = {https://doi.org/10.1038/299802a0}
}

@article{2023nrpPhoton,
  author = {Couteau, Christophe and others},
  title = {Applications of Single Photons to Quantum Communication and
    Computing},
  journal = {Nat. Rev. Phys.},
  volume = {5},
  number = {6},
  pages = {326-338},
  year = {2023},
  url = {https://doi.org/10.1038/s42254-023-00583-2}
}

@article{cz1998repeater,
  title = {Quantum Repeaters: The Role of Imperfect Local Operations in Quantum Communication},
  author = {Briegel, H.-J. and D\"ur, W. and Cirac, J. I. and Zoller, P.},
  journal = {Phys. Rev. Lett.},
  volume = {81},
  issue = {26},
  pages = {5932--5935},
  numpages = {0},
  year = {1998},
  month = {Dec},
  publisher = {American Physical Society},
  doi = {10.1103/PhysRevLett.81.5932},
  url = {https://link.aps.org/doi/10.1103/PhysRevLett.81.5932},
}

@article{2024QDEntangle,
  author = {Laccotripes, P. and others},
  title = {Spin-Photon Entanglement with Direct Photon Emission in the
    Telecom {C-band}},
  journal = {Nat. Commun.},
  volume = {15},
  number = {1},
  pages = {9740},
  year = {2024},
  url = {https://doi.org/10.1038/s41467-024-53964-1}
}

@article{reichel_nondegenerate,
doi = {10.1088/1367-2630/15/4/045002},
url = {https://doi.org/10.1088/1367-2630/15/4/045002},
year = {2013},
month = {apr},
publisher = {IOP Publishing},
volume = {15},
number = {4},
pages = {045002},
author = {Miguel-Sánchez, Javier and others},
title = {Cavity quantum electrodynamics with charge-controlled quantum dots coupled to a fiber Fabry–Perot cavity},
journal = {New J. Phys.},
}

@article{zhoukun_nondegenerate,
    author = {Cui, Jin-Ming and others},
    title = {Polarization nondegenerate fiber Fabry-Perot cavities with large tunable splittings},
    journal = {Appl. Phys. Lett.},
    volume = {112},
    number = {17},
    pages = {171105},
    year = {2018},
    month = {04},
    issn = {0003-6951},
    doi = {10.1063/1.5024798},
    url = {https://doi.org/10.1063/1.5024798},
}

@Article{Reichel_cavity_bec,
author={Colombe, Yves
and others},
title={Strong atom--field coupling for Bose--Einstein condensates in an optical cavity on a chip},
journal={Nature},
year={2007},
month={Nov},
day={01},
volume={450},
number={7167},
pages={272-276},
issn={1476-4687},
doi={10.1038/nature06331},
url={https://doi.org/10.1038/nature06331}
}

@Article{qd_teleemission_shields,
author={M{\"u}ller, T.
and others},
title={A quantum light-emitting diode for the standard telecom window around 1,550{\thinspace}nm},
journal={Nat. Commun.},
year={2018},
month={Feb},
day={28},
volume={9},
number={1},
pages={862},
issn={2041-1723},
doi={10.1038/s41467-018-03251-7},
url={https://doi.org/10.1038/s41467-018-03251-7}
}

@article{2025ErEntangle,
  title = {Spin-Photon Entanglement of a Single ${\mathrm{Er}}^{3+}$ Ion in the Telecom Band},
  author = {Uysal, Mehmet T. and others},
  journal = {Phys. Rev. X},
  volume = {15},
  issue = {1},
  pages = {011071},
  numpages = {18},
  year = {2025},
  month = {Mar},
  publisher = {American Physical Society},
  doi = {10.1103/PhysRevX.15.011071},
  url = {https://link.aps.org/doi/10.1103/PhysRevX.15.011071}
}

@article{er_reiserer_stable,
author = {Alexander Ulanowski  and Benjamin Merkel  and Andreas Reiserer },
title = {Spectral multiplexing of telecom emitters with stable transition frequency},
journal = {Sci. Adv.},
volume = {8},
number = {43},
pages = {eabo4538},
year = {2022},
doi = {10.1126/sciadv.abo4538},
URL = {https://www.science.org/doi/abs/10.1126/sciadv.abo4538},}

@article{Er_indis_thomp,
  author = {Ourari, Salim and others},
  title = {Indistinguishable Telecom Band Photons from a Single {Er} Ion
    in the Solid State},
  journal = {Nature},
  volume = {620},
  number = {7976},
  pages = {977-981},
  year = {2023},
  url = {https://doi.org/10.1038/s41586-023-06281-4}
}

@article{hanson2018review,
author = {S. Wehner  and D. Elkouss  and R. Hanson },
title = {Quantum internet: A vision for the road ahead},
journal = {Science},
volume = {362},
number = {6412},
pages = {eaam9288},
year = {2018},
doi = {10.1126/science.aam9288},
URL = {https://www.science.org/doi/abs/10.1126/science.aam9288},
}

@article{wangshuangQKD,
  author = {Wang, Shuang and others},
  title = {Twin-Field Quantum Key Distribution over 830-Km Fibre},
  journal = {Nat. Photon.},
  volume = {16},
  number = {2},
  pages = {154-161},
  year = {2022},
  url = {https://doi.org/10.1038/s41566-021-00928-2}
}

@article{panQKD,
  title = {Experimental Twin-Field Quantum Key Distribution over 1000 km Fiber Distance},
  author = {Liu, Yang and others},
  journal = {Phys. Rev. Lett.},
  volume = {130},
  issue = {21},
  pages = {210801},
  numpages = {6},
  year = {2023},
  month = {May},
  publisher = {American Physical Society},
  doi = {10.1103/PhysRevLett.130.210801},
  url = {https://link.aps.org/doi/10.1103/PhysRevLett.130.210801}
}

@article{zhouErstorage,
  title = {On-Demand Storage of Photonic Qubits at Telecom Wavelengths},
  author = {Liu, Duan-Cheng and others},
  journal = {Phys. Rev. Lett.},
  volume = {129},
  issue = {21},
  pages = {210501},
  numpages = {6},
  year = {2022},
  month = {Nov},
  publisher = {American Physical Society},
  doi = {10.1103/PhysRevLett.129.210501},
  url = {https://link.aps.org/doi/10.1103/PhysRevLett.129.210501}
}

@Article{lukin2024qec,
author={Bluvstein, Dolev
and others},
title={Logical quantum processor based on reconfigurable atom arrays},
journal={Nature},
year={2024},
month={Feb},
day={01},
volume={626},
number={7997},
pages={58-65},
issn={1476-4687},
doi={10.1038/s41586-023-06927-3},
url={https://doi.org/10.1038/s41586-023-06927-3}
}

@article{endres6100,
  author = {Manetsch, Hannah J. and others},
  title = {A Tweezer Array with 6,100 Highly Coherent Atomic Qubits},
  journal = {Nature},
  volume = {647},
  number = {8088},
  pages = {60-67},
  year = {2025},
  url = {https://doi.org/10.1038/s41586-025-09641-4}
}

@Article{bernien2023npj,
author={Covey, Jacob P.
and Weinfurter, Harald
and Bernien, Hannes},
title={Quantum networks with neutral atom processing nodes},
journal={npj Quantum Inf.},
year={2023},
month={Sep},
day={16},
volume={9},
number={1},
pages={90},
issn={2056-6387},
doi={10.1038/s41534-023-00759-9},
url={https://doi.org/10.1038/s41534-023-00759-9}
}

@article{rempe2015rmp,
  title = {Cavity-based quantum networks with single atoms and optical photons},
  author = {Reiserer, Andreas and Rempe, Gerhard},
  journal = {Rev. Mod. Phys.},
  volume = {87},
  issue = {4},
  pages = {1379--1418},
  numpages = {40},
  year = {2015},
  month = {Dec},
  publisher = {American Physical Society},
  doi = {10.1103/RevModPhys.87.1379},
  url = {https://link.aps.org/doi/10.1103/RevModPhys.87.1379}
}

@article{saffmanqc,
  title = {Universal Neutral-Atom Quantum Computer with Individual Optical Addressing and Nondestructive Readout},
  author = {Radnaev, A.G. and others},
  journal = {PRX Quantum},
  volume = {6},
  issue = {3},
  pages = {030334},
  numpages = {20},
  year = {2025},
  month = {Aug},
  publisher = {American Physical Society},
  doi = {10.1103/66s8-jj18},
  url = {https://link.aps.org/doi/10.1103/66s8-jj18}
}

@phdthesis{chen2022thesis_rydber_ring,
  title={A Platform for Cavity Quantum Electrodynamics with Rydberg Atom Arrays},
  author={Chen, Yu-Ting},
  year={2022},
  school={Harvard University}
}

@phdthesis{kong2018thesis_waterloo,
  title={Towards many-body physics with Rydberg-dressed cavity polaritons},
  author={Kong, Hyeran},
  year={2018},
  school={University of Waterloo}
}

@misc{britton_zno_cavity,
      title={Can TCOs Transform Cavity-QED?}, 
      author={Wance Wang and Dhruv Fomra and Amit Agrawal and Henri J. Lezec and Joseph W. Britton},
      year={2025},
      archivePrefix={arXiv},
      primaryClass={quant-ph},
      url={https://arxiv.org/abs/2506.02501}, 
}

@article{cui_fiber_metalize_2026,
  title={Design and fabrication of metal-shielded fiber-cavity mirrors for ion-trap systems},
  author={Chen, Wei-Bin and others},
  journal={Quantum Science and Technology},
  volume={11},
  number={1},
  pages={015045},
  year={2026},
  publisher={IOP Publishing}
}

@Article{lukin_2026_faulttolerant,
author={Bluvstein, Dolev
and others},
title={A fault-tolerant neutral-atom architecture for universal quantum computation},
journal={Nature},
year={2026},
month={Jan},
day={01},
volume={649},
number={8095},
pages={39-46},
issn={1476-4687},
doi={10.1038/s41586-025-09848-5},
url={https://doi.org/10.1038/s41586-025-09848-5}
}

@article{northup_ion_ffpc_2023,
    author = {Teller, Markus and others},
    title = {Integrating a fiber cavity into a wheel trap for strong ion–cavity coupling},
    journal = {AVS Quantum Science},
    volume = {5},
    number = {1},
    pages = {012001},
    year = {2023},
    month = {01},
    issn = {2639-0213},
    doi = {10.1116/5.0121534},
    url = {https://doi.org/10.1116/5.0121534},
}

@article{saffman_parabolic_prxq2026,
  title = {Efficient and Compact Quantum Network Node Based on a Parabolic Mirror on an Optical Chip},
  author = {Safari, A. and Oh, E. and Huft, P. and Chase, G. and Zhang, J. and Saffman, M.},
  journal = {PRX Quantum},
  volume = {7},
  issue = {3},
  pages = {033008},
  numpages = {18},
  year = {2026},
  month = {Jul},
  publisher = {American Physical Society},
  doi = {10.1103/fhf3-3nzb},
  url = {https://link.aps.org/doi/10.1103/fhf3-3nzb}
}

@article{rempe_xcav_776,
  title = {Source of Heralded Atom-Photon Entanglement for Quantum Networking},
  author = {Chiarella, Gianvito and Frank, Tobias and Zuka, Leart and Farrera, Pau and Rempe, Gerhard},
  journal = {Phys. Rev. Lett.},
  volume = {135},
  issue = {24},
  pages = {240802},
  numpages = {7},
  year = {2025},
  month = {Dec},
  publisher = {American Physical Society},
  doi = {10.1103/5zk9-3rpv},
  url = {https://link.aps.org/doi/10.1103/5zk9-3rpv}
}

@article{weinfurter_2026_metropolitan,
  title = {Metropolitan entanglement distribution between an atom and a near-visible photon},
  author = {Büki, Maya and Malik, Pooja and Fertig, Florian and Frank, Tobias and Scholz, Marvin and Block, Tommy and Chiarella, Gianvito and Zhou, Yiru and Distante, Emanuele and Farrera, Pau and Rempe, Gerhard and Weinfurter, Harald},
  journal = {Phys. Rev. Lett.},
  pages = {},
  year = {2026},
  month = {Jun},
  publisher = {American Physical Society},
  doi = {10.1103/94hz-xtht},
  url = {https://link.aps.org/doi/10.1103/94hz-xtht}
}

@article{weinfurter_qfc_2020,
  title = {Long-Distance Distribution of Atom-Photon Entanglement at Telecom Wavelength},
  author = {van Leent, Tim and others},
  journal = {Phys. Rev. Lett.},
  volume = {124},
  issue = {1},
  pages = {010510},
  numpages = {6},
  year = {2020},
  month = {Jan},
  publisher = {American Physical Society},
  doi = {10.1103/PhysRevLett.124.010510},
  url = {https://link.aps.org/doi/10.1103/PhysRevLett.124.010510}
}

@article{33km,
  author = {Leent, Tim van and others},
  title = {Entangling Single Atoms over 33 Km Telecom Fibre},
  journal = {Nature},
  volume = {607},
  number = {7917},
  pages = {69-73},
  year = {2022},
  url = {https://doi.org/10.1038/s41586-022-04764-4}
}

@article{covey_clock_network,
  title = {Probing Curved Spacetime with a Distributed Atomic Processor Clock},
  author = {Covey, Jacob P. and Pikovski, Igor and Borregaard, Johannes},
  journal = {PRX Quantum},
  volume = {6},
  issue = {3},
  pages = {030310},
  numpages = {12},
  year = {2025},
  month = {Jul},
  publisher = {American Physical Society},
  doi = {10.1103/q188-b1cr},
  url = {https://link.aps.org/doi/10.1103/q188-b1cr}
}

@article{covey_multiplexing_cavity,
  title = {Multiplexed telecommunication-band quantum networking with atom arrays in optical cavities},
  author = {Huie, William and Menon, Shankar G. and Bernien, Hannes and Covey, Jacob P.},
  journal = {Phys. Rev. Res.},
  volume = {3},
  issue = {4},
  pages = {043154},
  numpages = {14},
  year = {2021},
  month = {Dec},
  publisher = {American Physical Society},
  doi = {10.1103/PhysRevResearch.3.043154},
  url = {https://link.aps.org/doi/10.1103/PhysRevResearch.3.043154}
}

@article{kuzmich_telecom,
  title = {Quantum Telecommunication Based on Atomic Cascade Transitions},
  author = {Chaneli\`ere, T. and Matsukevich, D. N. and Jenkins, S. D. and Kennedy, T. A. B. and Chapman, M. S. and Kuzmich, A.},
  journal = {Phys. Rev. Lett.},
  volume = {96},
  issue = {9},
  pages = {093604},
  numpages = {4},
  year = {2006},
  month = {Mar},
  publisher = {American Physical Society},
  doi = {10.1103/PhysRevLett.96.093604},
  url = {https://link.aps.org/doi/10.1103/PhysRevLett.96.093604}
}

@article{2025ybentangle,
  author = {Li, Lintao and others},
  title = {Parallelized Telecom Quantum Networking with an Ytterbium-171
    Atom Array},
  journal = {Nat. Phys.},
  volume = {21},
  number = {11},
  pages = {1826-1833},
  year = {2025},
  url = {https://doi.org/10.1038/s41567-025-03022-4}
}

@article{duan1530,
  title = {Long-Distance Entanglement between a Multiplexed Quantum Memory and a Telecom Photon},
  author = {Chang, W. and others},
  journal = {Phys. Rev. X},
  volume = {9},
  issue = {4},
  pages = {041033},
  numpages = {9},
  year = {2019},
  month = {Nov},
  publisher = {American Physical Society},
  doi = {10.1103/PhysRevX.9.041033},
  url = {https://link.aps.org/doi/10.1103/PhysRevX.9.041033}
}

@article{2014clocknetwork,
  author = {Kómár, P. and others},
  title = {A Quantum Network of Clocks},
  journal = {Nature Physics},
  volume = {10},
  number = {8},
  pages = {582-587},
  year = {2014},
  url = {https://doi.org/10.1038/nphys3000}
}

@article{
lukinFibercav,
author = {Brandon Grinkemeyer  and others},
title = {Error-detected quantum operations with neutral atoms mediated by an optical cavity},
journal = {Science},
volume = {387},
number = {6740},
pages = {1301-1305},
year = {2025},
doi = {10.1126/science.adr7075},
URL = {https://www.science.org/doi/abs/10.1126/science.adr7075},
}

@article{2025Readout,
  title = {Ultrafast High-Fidelity State Readout of Single Neutral Atom},
  author = {Wang, Jian and others},
  journal = {Phys. Rev. Lett.},
  volume = {134},
  issue = {24},
  pages = {240802},
  numpages = {7},
  year = {2025},
  month = {Jun},
  publisher = {American Physical Society},
  doi = {10.1103/PhysRevLett.134.240802},
  url = {https://link.aps.org/doi/10.1103/PhysRevLett.134.240802}
}

@article{WeinIonization,
  title = {Event-Ready Bell Test Using Entangled Atoms Simultaneously Closing Detection and Locality Loopholes},
  author = {Rosenfeld, Wenjamin and others},
  journal = {Phys. Rev. Lett.},
  volume = {119},
  issue = {1},
  pages = {010402},
  numpages = {6},
  year = {2017},
  month = {Jul},
  publisher = {American Physical Society},
  doi = {10.1103/PhysRevLett.119.010402},
  url = {https://link.aps.org/doi/10.1103/PhysRevLett.119.010402}
}

@article{
Wein2012Entangle,
author = {Julian Hofmann and others},
title = {Heralded Entanglement Between Widely Separated Atoms},
journal = {Science},
volume = {337},
number = {6090},
pages = {72-75},
year = {2012},
doi = {10.1126/science.1221856},
URL = {https://www.science.org/doi/abs/10.1126/science.1221856},
}

@article{RempeXcav,
  author = {Brekenfeld, Manuel and Niemietz, Dominik and Christesen,
    Joseph Dale and Rempe, Gerhard},
  title = {A Quantum Network Node with Crossed Optical Fibre Cavities},
  journal = {Nature Physics},
  volume = {16},
  number = {6},
  pages = {647-651},
  year = {2020},
  url = {https://doi.org/10.1038/s41567-020-0855-3}
}

@article{pandiqkd,
author = {Bo-Wei Lu  and others },
title = {Device-independent quantum key distribution over 100 km with single atoms},
journal = {Science},
volume = {391},
number = {6785},
pages = {592-597},
year = {2026},
doi = {10.1126/science.aec6243},
URL = {https://www.science.org/doi/abs/10.1126/science.aec6243},
}

@article{monroe2004,
  author = {Blinov, B. B. and Moehring, D. L. and Duan, L.- M. and
    Monroe, C.},
  title = {Observation of Entanglement Between a Single Trapped Atom and
    a Single Photon},
  journal = {Nature},
  volume = {428},
  number = {6979},
  pages = {153-157},
  year = {2004},
  url = {https://doi.org/10.1038/nature02377}
}

@article{HansonLoophole,
  author = {Storz, Simon and others},
  title = {Loophole-Free {Bell} Inequality Violation with
    Superconducting Circuits},
  journal = {Nature},
  volume = {617},
  number = {7960},
  pages = {265-270},
  year = {2023},
  url = {https://doi.org/10.1038/s41586-023-05885-0}
}

@article{xiaolong2photon,
  title = {Purcell-Enhanced Generation of Photonic Bell States via the Inelastic Scattering off Single Atoms},
  author = {Wang, Jian and others},
  journal = {Phys. Rev. Lett.},
  volume = {134},
  issue = {5},
  pages = {053401},
  numpages = {7},
  year = {2025},
  month = {Feb},
  publisher = {American Physical Society},
  doi = {10.1103/PhysRevLett.134.053401},
  url = {https://link.aps.org/doi/10.1103/PhysRevLett.134.053401}
}

@article{rempe100ms,
  author = {Körber, M. and others},
  title = {Decoherence-Protected Memory for a Single-Photon Qubit},
  journal = {Nat. Photon.},
  volume = {12},
  number = {1},
  pages = {18-21},
  year = {2018},
  url = {https://doi.org/10.1038/s41566-017-0050-y}
}

@article{rempeTelePropos,
  author = {Uphoff, Manuel and Brekenfeld, Manuel and Rempe, Gerhard and
    Ritter, Stephan},
  title = {An Integrated Quantum Repeater at Telecom Wavelength with
    Single Atoms in Optical Fiber Cavities},
  journal = {Appl. Phys. B},
  volume = {122},
  number = {3},
  pages = {46},
  year = {2016},
  url = {https://doi.org/10.1007/s00340-015-6299-2}
}

@article{BernienTelePropos,
doi = {10.1088/1367-2630/ab98d4},
url = {https://doi.org/10.1088/1367-2630/ab98d4},
year = {2020},
month = {jul},
publisher = {IOP Publishing},
volume = {22},
number = {7},
pages = {073033},
author = {Menon, Shankar G and Singh, Kevin and Borregaard, Johannes and Bernien, Hannes},
title = {Nanophotonic quantum network node with neutral atoms and an integrated telecom interface},
journal = {New J. Phys.},
}

@article{CoveyTelePropos,
  title = {Telecom-Band Quantum Optics with Ytterbium Atoms and Silicon Nanophotonics},
  author = {Covey, Jacob P. and others},
  journal = {Phys. Rev. Appl.},
  volume = {11},
  issue = {3},
  pages = {034044},
  numpages = {15},
  year = {2019},
  month = {Mar},
  publisher = {American Physical Society},
  doi = {10.1103/PhysRevApplied.11.034044},
  url = {https://link.aps.org/doi/10.1103/PhysRevApplied.11.034044}
}

@article{qdTelecomReview,
  author = {Yu, Ying and others},
  title = {Telecom-Band Quantum Dot Technologies for Long-Distance
    Quantum Networks},
  journal = {Nat. Nanotechnol.},
  volume = {18},
  number = {12},
  pages = {1389-1400},
  year = {2023},
  url = {https://doi.org/10.1038/s41565-023-01528-7}
}

@misc{bands_knowledge,
    note = {Telecom bands in optical fibers span from 1250 nm to 1650 nm, with the C-band (1530–1565 nm) exhibiting the minimum loss. See \cite{sup} for more details.}
}

@misc{sup, 
      note={See supplemental material for additional information about entanglement generation effciency, fidelity estimation, analysis of different wavelength in entanglement generation, and future scalability and compatibility with Rydberg-mediated gates.}
}

@Article{qd_indis_syperek,
author={Holewa, Pawe{\l}
and others},
title={High-throughput quantum photonic devices emitting indistinguishable photons in the telecom C-band},
journal={Nat. Commun.},
year={2024},
month={Apr},
day={18},
volume={15},
number={1},
pages={3358},
issn={2041-1723},
doi={10.1038/s41467-024-47551-7},
url={https://doi.org/10.1038/s41467-024-47551-7}
}

@Article{qd_indis_hoefling ,
author={Hauser, Nico
and others},
title={Deterministic and highly indistinguishable single photons in the telecom C-band},
journal={Nat. Commun.},
year={2026},
month={Jan},
day={14},
volume={17},
number={1},
pages={537},
issn={2041-1723},
doi={10.1038/s41467-026-68336-0},
url={https://doi.org/10.1038/s41467-026-68336-0}
}

@article{er_groeblacher,
  title = {Frequency Tunable, Cavity-Enhanced Single Erbium Quantum Emitter in the Telecom Band},
  author = {Yu, Yong and others},
  journal = {Phys. Rev. Lett.},
  volume = {131},
  issue = {17},
  pages = {170801},
  numpages = {7},
  year = {2023},
  month = {Oct},
  publisher = {American Physical Society},
  doi = {10.1103/PhysRevLett.131.170801},
  url = {https://link.aps.org/doi/10.1103/PhysRevLett.131.170801}
}

@Article{er_hxtang,
author={Yang, Likai
and Wang, Sihao
and Shen, Mohan
and Xie, Jiacheng
and Tang, Hong X.},
title={Controlling single rare earth ion emission in an electro-optical nanocavity},
journal={Nat. Commun.},
year={2023},
month={Mar},
day={28},
volume={14},
number={1},
pages={1718},
issn={2041-1723},
doi={10.1038/s41467-023-37513-w},
url={https://doi.org/10.1038/s41467-023-37513-w}
}

@Article{qd_telecom_htoon,
author={Zhao, Huan
and Pettes, Michael T.
and Zheng, Yu
and Htoon, Han},
title={Site-controlled telecom-wavelength single-photon emitters in atomically-thin MoTe2},
journal={Nat. Commun.},
year={2021},
month={Nov},
day={19},
volume={12},
number={1},
pages={6753},
issn={2041-1723},
doi={10.1038/s41467-021-27033-w},
url={https://doi.org/10.1038/s41467-021-27033-w}
}

@book{fund_photonics,
    author = {Bahaa E. A. Saleh, Malvin Carl Teich},
    title = {Fundamentals of Photonics},
    publisher = {Wiley},
    year = {2020}
}

@article{riedmatten_er_purcell,
  author = {Casabone, Bernardo and others},
  title = {Dynamic Control of {Purcell} Enhanced Emission of Erbium Ions
    in Nanoparticles},
  journal = {Nature Communications},
  volume = {12},
  number = {1},
  pages = {3570},
  year = {2021},
  url = {https://doi.org/10.1038/s41467-021-23632-9}
}

@article{xiaolongpid,
title = {Active stabilization of multi-parameter in AMO experiments with a single digital servo},
journal = {Opt. Laser Technol.
},
volume = {167},
pages = {109791},
year = {2023},
issn = {0030-3992},
doi = {https://doi.org/10.1016/j.optlastec.2023.109791},
url = {https://www.sciencedirect.com/science/article/pii/S0030399223006849},
author = {Xiao-Long Zhou and others},
}

@article{shenRSI,
    author = {Shen, Ze-Min and others},
    title = {Continuously and widely tunable frequency-stabilized laser based on an optical frequency comb},
    journal = {Rev. Sci. Instrum.},
    volume = {94},
    number = {2},
    pages = {023001},
    year = {2023},
    month = {02},
    issn = {0034-6748},
    doi = {10.1063/5.0120119},
    url = {https://doi.org/10.1063/5.0120119},
}

\section*{Methods}

\setcounter{figure}{0}
\renewcommand{\figurename}{Extended Data Fig.}
\setcounter{table}{0}
\renewcommand{\tablename}{Extended Data Table.}

\subsection*{Experimental setup}

\begin{figure*}[h]
    \centering
    \includegraphics[width=\textwidth]{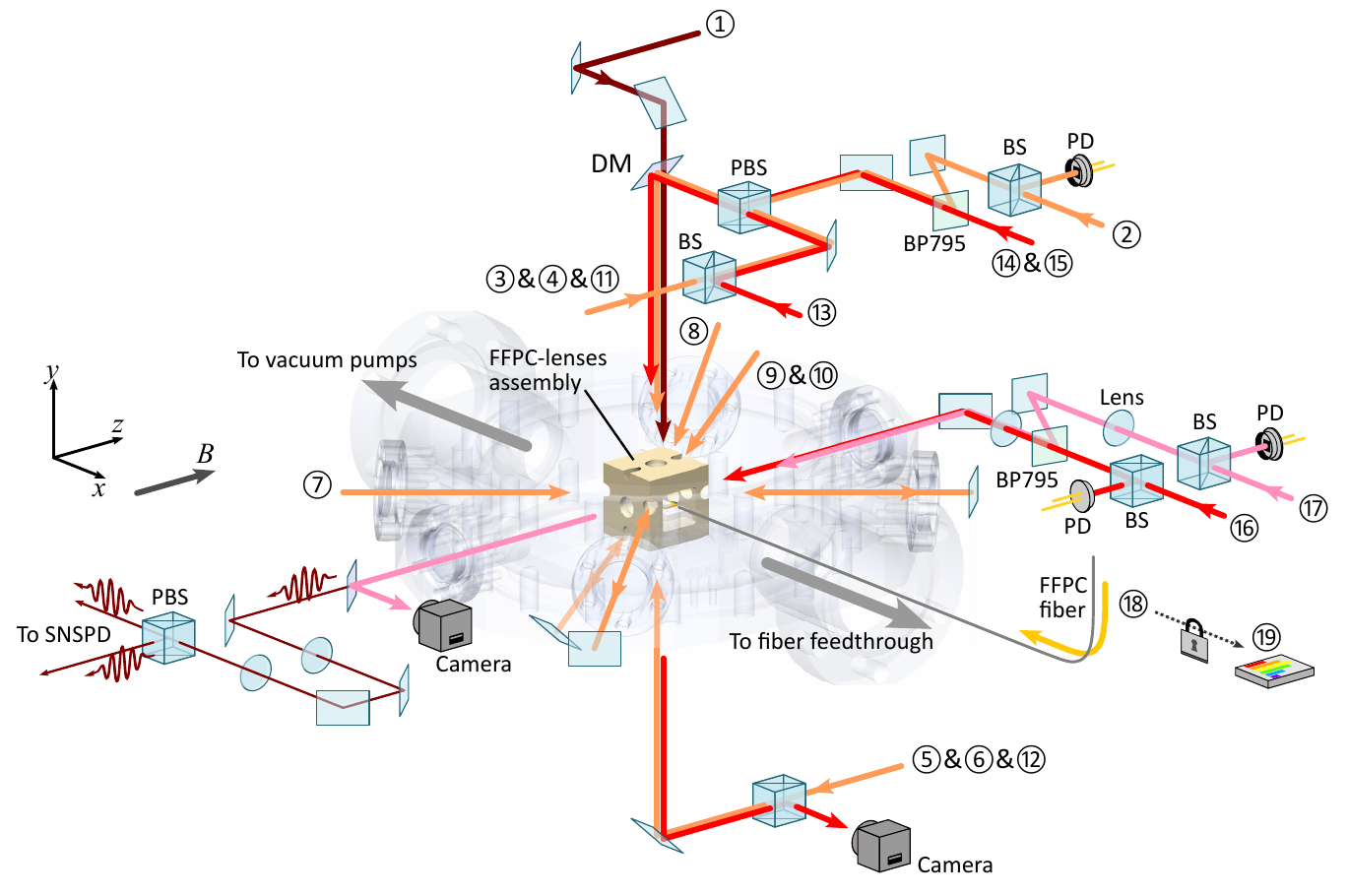}
    \caption{An overview of the experimental setup. Lying at the center of a vacuum chamber is an assembly consisting of an FFPC and four aspherical lenses glued to a PEEK-machined base. The corresponding lasers with their functions are listed in \exttab{Et1}. Not all the optical elements are shown. DM: dichroic mirror. BS: beam splitter. PBS: polarization beam splitter. PD: photodetector (for power stabilization). BP: band-pass filter.}
    \label{E1}
\end{figure*}

An overview of the experimental setup is given in \extfig{E1} and \exttab{Et1}. The FFPC is integrated with four aspherical lenses (LightPath 355397) placed inside the vacuum chamber. The lenses serve as the optical channels for cooling, trapping, state operation and photon collection of the atom.

\begin{figure*}
    \centering
    \includegraphics[width=0.6\textwidth]{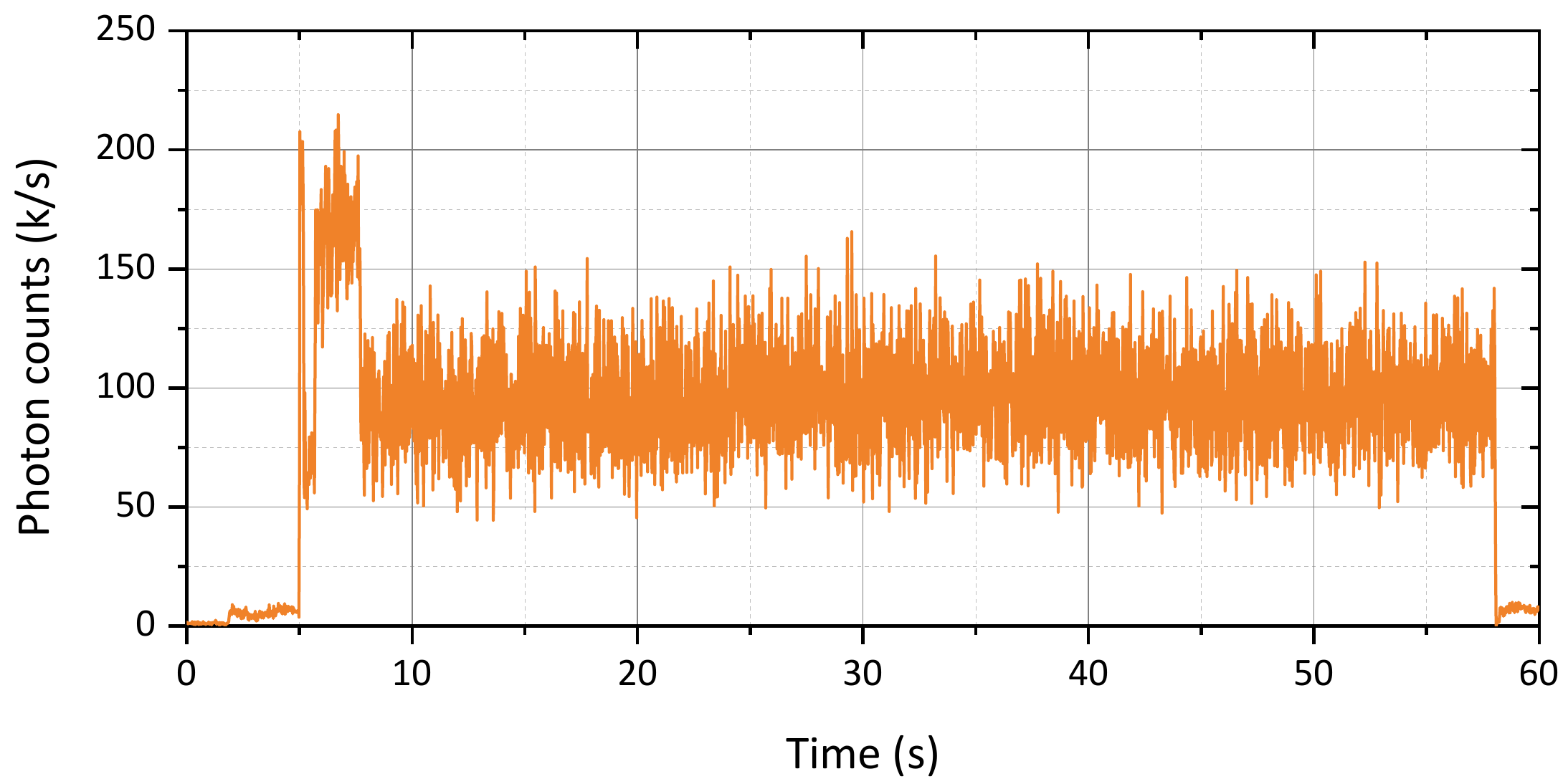}
    \caption{Photon counts from the out-coupling of FFPC during the cooling of a single atom in a typical run. The integration bin of photon counts is 20 ms. A decrease at approximately 7.5 s results from the decrease in the duty ratio of laser cooling during the experiment.}
    \label{E2}
\end{figure*}

Our experiment begins with preparing a magneto-optical trap (MOT) which is 2 mm above the fiber cavity. After releasing the MOT by shutting down the magnetic gradient and cooling lasers, a near-resonant 795 nm laser beam serves as a guiding potential to guide the free-falling atoms to the center of the cavity. Intra-cavity laser cooling is performed subsequently, as described in our previous work for single atom loading in an optical lattice \cite{2025readout}. Here, the vertical optical lattice in \cite{2025readout} is replaced by a horizontal-propagating optical tweezer. The reason is that in our previous work, the beam waist of the optical tweezer is more than 10 $\mu$m, which results in the uncertainty of the atom's position along the cavity axis. Therefore, we apply an 850 nm optical tweezer instead to capture the laser-cooled atom from the guiding potential, and pinpoint the single atom on the focal plane of the aspherical lenses. Under continuous laser cooling of the atom, the scattered photons are collected and out-coupled from the FFPC, resulting in the distinct fluorescence signals detected by the single photon detector as shown in \extfig{E2}.

The free-space telecom photon collection requires precise and stable alignment of the collection mode and the single atom. As shown in Fig. \extfig{E1}, the optical tweezer shares the same beam path with the 1530 nm collection mode. Therefore, the coarse alignment can be achieved by the mutual coupling of two single mode fibers. However, the aspherical lenses (LightPath 355397) are not designed to work in the telecom C-band, which causes dramatic chromatic aberration between 850 nm and 1530 nm. To further overlap the two laser modes, we send the 1530 nm excitation laser in the collection fiber to drive the $5S$--$5P$--$4D$ two-photon transition together with the original 780 nm excitation laser. By maximizing the photon counts from the cascaded decay into the fiber cavity, we optimize the overlap between the atom and the beam waist of the collection mode.

\subsection*{Magnetic Field Stabilization}
\begin{figure*}
	\centering
	\includegraphics[width=0.85\textwidth]{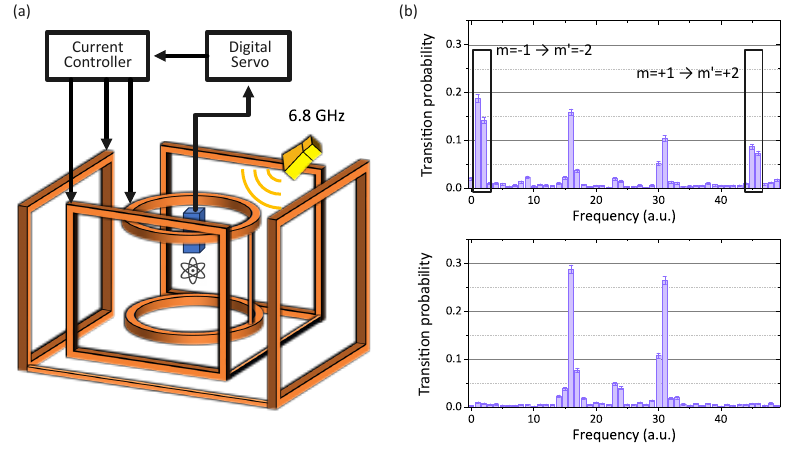}
	\caption{(a) Experimental setup for magnetic field stabilization. The magnetic sensor is placed close to the vacuum chamber, whose voltages are fed into the digital servo to drive the three Helmholtz coils. A microwave horn is applied for driving magnetic dipole transitions between the hyperfine levels. (b) Microwave spectrum of the $F=1$ level before (up) and after (down) state initialization. The magnetic field is tuned such that the transition probability is minimized when the microwave frequency is resonant with transitions $m=-1\rightarrow m=-2$ and $m=+1\rightarrow m=+2$ as marked in the plot.}
	\label{E3}
\end{figure*}

In our experiment, a stabilized magnetic field is required for a well-defined quantization axis and maintaining the coherence of the atomic qubit. Here, we apply a bias field of 170 mG along the $z$-axis as shown in the main text. The magnetic field is generated by three independent pairs of Helmholtz coils whose current is supplied by three channels of precise current controllers (Vescent, ICE-DCC-500) with external modulation ports and measured by a magnetic sensor (Honor Top, HT-191) outside the vacuum chamber. Therefore, we use an FPGA-based multi-parameter digital servo \cite{xiaolongpid} to simultaneously stabilize the magnetic field along three axes to constant values as shown in \extfig{E3} (a). As shown in the time sequence in the main text, the magnetic field stabilization is only activated by triggering the controller after we shut down the MOT whose large magnetic gradient field can easily saturate the sensor.\\

To ensure that the magnetic field is well aligned with the polarization of the $\pi$-polarized lasers, we optically pump the atoms to the state $\ket{5^2S_{1/2},F=1,m=0}$ and measure the relative population distribution in the $F=1$ level using microwave transitions. After the optical pumping, we apply a $50$-$\mu$s long microwave pulse with varied driving frequency $\omega$ and measure the population in $F=2$. A transition between hyperfine levels is only possible when $\omega$ equals a transition between two Zeeman states in two hyperfine manifolds. Therefore, we can tune the setpoint of the digital servo by minimizing the transitions from $\ket{5^2S_{1/2},F=1,m=\pm 1}$ to $F=2$, with a result shown in \extfig{E3} (b).

\subsection*{Two-photon Excitation}

\subsubsection*{Laser Setup}

\begin{figure*}
	\centering
	\includegraphics[width=0.6\textwidth]{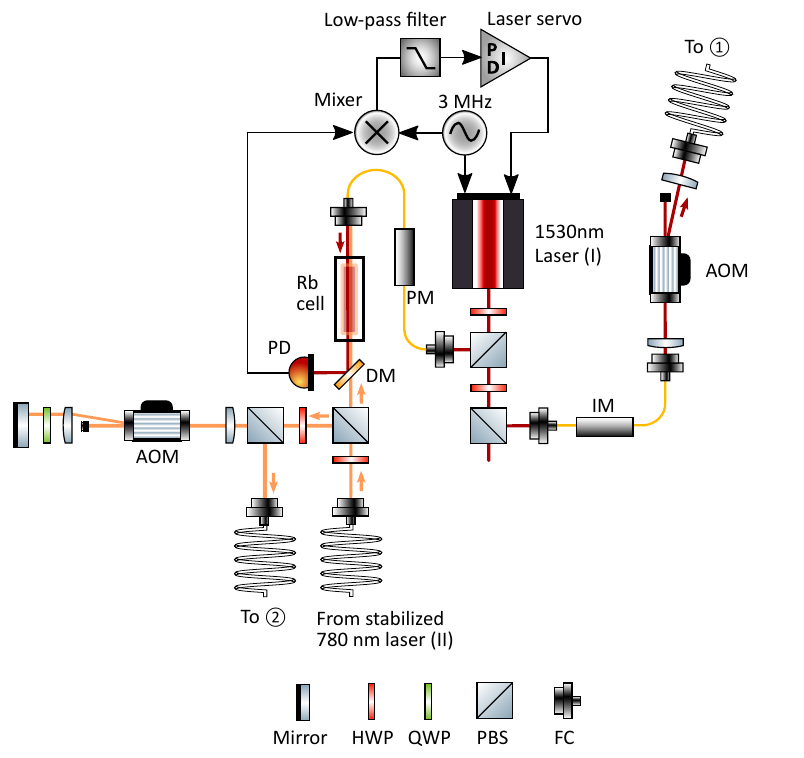}
	\caption{Frequency stabilization and switching of the excitation lasers. The 1530 nm diode laser is stabilized on the two-photon resonant absorption line $5^2S_{1/2},F=2$ to $4^2D_{5/2},F=4$ in the heated rubidium-87 vapor cell, with the 780 nm laser stabilized on the cross-over signal between $5^2S_{1/2},F=2$ to $5^2P_{3/2},F=2$ and $5^2S_{1/2},F=2$ to $5^2P_{3/2},F=3$ (not shown). The locking point is adjusted by tuning the frequency of the electro-optic phase modulator. After passing through the switching AOMs, the two laser beams achieve two-photon resonance with the target transition. PD: photodetector. DM: dichroic mirror. IM: electro-optic intensity modulator. PM: electro-optic phase modulator. HWP (QWP): half (quarter) wave-plate. PBS: polarization beam splitter. FC: fiber collimator.}
	\label{E4}
\end{figure*}

To coherently drive the atom from the ground state $\ket{g}$ to the excited state $\ket{e}$, the excitation lasers have to maintain resonance with the cascaded transition. With our 780 nm laser stabilized by regular saturated absorption spectroscopy technology for laser cooling and state readout, we use two-photon resonance spectroscopy to stabilize the frequency of the 1530 nm laser as shown in \extfig{E4}. Prior to entering the rubidium vapor cell, a fiber electro-optic modulator for phase modulation generates frequency sidebands of the 1530 nm laser. The modulation frequency is chosen to be approximately 1 GHz, with the 780 nm laser passing through an AOM twice, such that the two-photon resonance is achieved by the laser beams that finally drive the atom.

\subsubsection*{Suppression of Unwanted Channel}

As mentioned in the main text, an excitation with large Rabi frequency may cause unwanted population in the state $\ket{e'} := \ket{4^2D_{5/2},F=2,m=0}$ other than the target state $\ket{e}:=\ket{4^2D_{5/2},F=4,m=0}$. Here we show that by choosing a specific Rabi frequency, this unwanted channel can be suppressed.

Given certain laser intensities, the Rabi frequency of a transition is proportional to the transition matrix element, which can be reduced and factorized using Wigner-Eckart theorem:
\[ \bra{F,m_F} er_q \ket{F',m_{F}'} = \bra{F}|e\bm{r}|\ket{F'}\bra{F,m_F}F',m_F';1,q\rangle, \]
where $q=0,\pm1$ denotes the polarization of the light field. The first term can be further simplified into
\begin{align*}
	\bra{F}|e\bm{r}|\ket{F'}
	= & \bra{J}|e\bm{r}|\ket{J'}(-1)^{F'+J+1+I} \\ &\sqrt{(2F'+1)(2J+1)} 
	\begin{Bmatrix}
		J & J' & 1 \\
		F' & F & I
	\end{Bmatrix},
\end{align*}
where $\{\cdot\}$ is the 6-$j$ symbol. Therefore, in our case, the relative amplitude of the two-photon Rabi frequencies of two transition channels:
\begin{align*}
	\mathrm{target: }\;\ket{g} &\rightarrow \ket{5^2P_{3/2},F=3,m=0} \rightarrow \ket{e}\\
	&\mathrm{and} \\
	\mathrm{unwanted: }\;\ket{g} &\rightarrow \ket{5^2P_{3/2},F=3,m=0} \rightarrow \ket{e'}\\
	& + \\
	\ket{g} &\rightarrow \ket{5^2P_{3/2},F=1,m=0} \rightarrow \ket{e'}\\
\end{align*}
can be derived. With a large single-photon detuning, we get
\[\frac{\Omega_{ge}}{\Omega_{ge'}} \approx 2.44\] when both transitions are on resonance.

With the frequency separation between $\ket{e}$ and $\ket{e'}$ $\Delta = 116$ MHz, the suppression of the unwanted transition requires the unwanted channel to go through a $2\pi$ rotation when we apply a $\pi-$pulse on the target channel, which demands the effective Rabi frequency
\[ \Omega_{ge'}' = \sqrt{\Delta^2 + \Omega_{ge'}^2} = 2\Omega_{ge} \]
and leads to a Rabi frequency $\Omega_{ge} = 60$ MHz.

\subsection*{Raman Laser Setup}

In our experiment, the Raman transfer protocol requires the two frequency components of the Raman laser to shine on the atom along different directions with respective polarizations. Therefore, regular modulation technique on laser phase or intensity is not feasible due to the complexity in narrow-band frequency filtering for beam separation. Also, the fast transfer process requires higher laser power, which can damage fiber-based modulators.

\begin{figure*}
	\centering
	\includegraphics[width=0.6\textwidth]{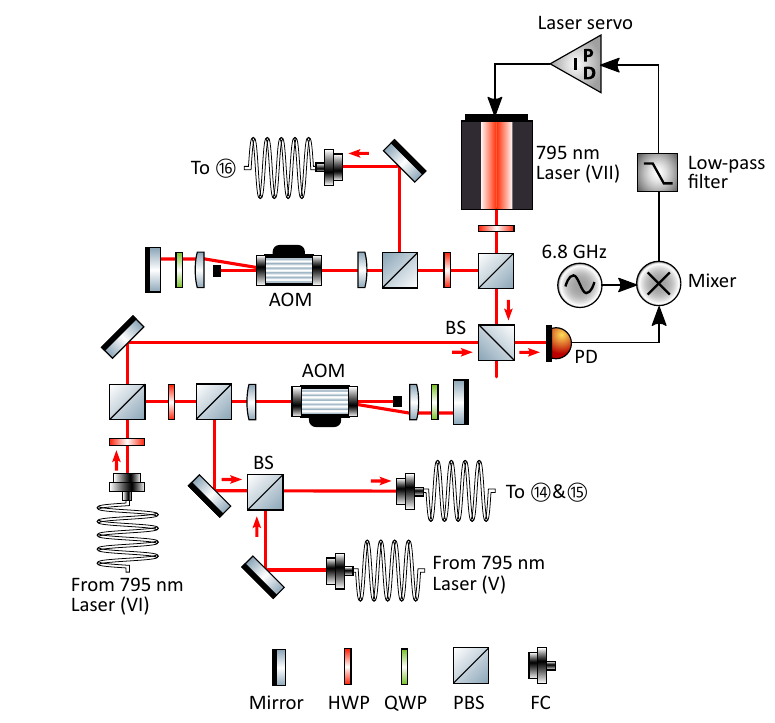}
	\caption{Experimental setup of Raman lasers. The two lasers are switched and sent to the atom separately. The beat signal generated by interference is mixed with a local microwave reference and then low-pass filtered to feed into an analog laser servo (Vescent D2-125). Therefore, with Laser (\Rmnum{6}) free-running, Laser (\Rmnum{7}) actively follows so that a stabilized frequency difference and coherence is maintained. PD: photodetector. DM: dichroic mirror. IM: electro-optic intensity modulator. PM: electro-optic phase modulator. HWP (QWP): half (quarter) wave-plate. PBS: polarization beam splitter. FC: fiber collimator.}
	\label{E5}
\end{figure*}

Here, we use two diode lasers (Moglabs CEL) with the frequency difference equal to the hyperfine splitting of the ground state $^{87}$Rb. The phase stability is ensured by applying optical phase-locked loop \cite{shenrsi}. The two lasers are sampled and interfered, with the beat signal compared with a local microwave reference to stabilize one of the Raman lasers as shown in \extfig{E5}. The digital servo mentioned above also serves for the power stabilization of the two Raman lasers, to eliminate long-term drift of the Rabi frequency which degrades the transfer efficiency.

\begin{table*}[h]
	\centering
	\caption{Lasers applied in the experiment. Indices of laser beams are corresponding to those shown in \extfig{E1}.}
    \setlength{\tabcolsep}{12pt}
	\begin{tabular}{l c l}
        \toprule
		Laser & Laser beam index & Function \\
		\midrule
		
		(\Rmnum{1}) Toptica DL pro 1530 
		& 1 & Resonant excitation \\
        \addlinespace
		
		(\Rmnum{2}) Precilaser 780 nm 
		& 2 & Resonant excitation \\
		& 3 \& 5 & Intra-cavity cooling \\
		& 4 \& 6 & Atomic state readout \\
		& 7, 8 \& 9 & MOT cooling \\
        \addlinespace
		
		(\Rmnum{3}) Uniquanta DFB 780 nm 
		& 10 & MOT repumping \\
		& 11 \& 12 & Intra-cavity repumping \\
        \addlinespace
		
		(\Rmnum{4}) Uniquanta DFB 795 nm 
		& 13 & Guiding of falling atoms \\
		
		(\Rmnum{5}) Uniquanta DFB 795 nm 
		& 14 & Optical pumping \\
		
		(\Rmnum{6}) Moglabs CEL 795 nm 
		& 15 & $\pi$-Raman transfer \\
		
		(\Rmnum{7}) Moglabs CEL 795 nm 
		& 16 & $\sigma$-Raman transfer \\
		
		(\Rmnum{8}) Moglabs ECDL 850 nm 
		& 17 & Optical tweezer \\
		
		(\Rmnum{9}) Moglabs CEL 776 nm 
		& 18 & Cavity stabilization \\
		
		(\Rmnum{10}) Menlo Systems FC1500-250-ULN
		& 19 & Laser frequency reference \\
		
        \bottomrule
	\end{tabular}
	\label{Et1}
\end{table*}

\vspace{+0.2cm}
\noindent
\textbf{Acknowledgement}

\noindent
This work is supported by National Key R\&D Program (No.\,2024YFA1409402), the Quantum Science and Technology-National Science and Technology Major Project (No.\,2021ZD0301200),  the National Natural Science Foundation of China (No.\,11821404), and the Fundamental Research Funds for the Central Universities (WK2470000038).
\newline

\noindent
\textbf{Author contributions}

\noindent
D.-Y. H. and J. W. designed and carried out the measurements, analyzed the data, and wrote the initial manuscript draft.
J. W. and C.-F. L. supervised key experimental developments and provided essential guidance on data interpretation.
G.-C. G. conceived the overall research direction, offered high-level scientific advice throughout the project.
Z.-M. S. and X.-L. Z. contributed to the theoretical modeling and numerical simulations.
S.-J. H. and Y.-J. L. supported data acquisition, performed auxiliary measurements, and contributed to figure preparation.
Q. J. and Y.-S. C. assisted in the construction and optimization of the laser setups.
\newline

\noindent
\textbf{Competing interests}

\noindent
The authors declare no competing interests.

\end{document}